# Modification of Turbulent Boundary Layer Coherent Structures with Drag Reducing Polymer Solution

Yasaman Farsiani, Zeeshan Saeed, Balaji Jayaraman & Brian R. Elbing[a]

Mechanical & Aerospace Engineering, Oklahoma State University, Stillwater, OK

## Abstract

The modification of dominant coherent structures that extend through the log-region of a drag reduced turbulent boundary layer is studied via examination of two-point correlations from time-resolved particle-image-velocimetry. Measurements were acquired in polymer oceans (uniform concentration) at drag reduction levels corresponding to the low drag reduction regime ($< 40\%$), the high drag reduction (HDR) regime ($>40\%$), and at an intermediate level ($46\%$). The mean velocity profiles and two-point correlations were compared with that of water (Newtonian, $DR = 0\%$). These results show that with increasing drag reduction the inclination of these dominant coherent structures decrease, their streamwise extent increases, and the fluctuations in the correlations are suppressed (especially at HDR). These observations are examined in comparison with the coherent structure literature (Newtonian and polymeric).

---

[a] Author to whom correspondence should be addressed. Electronic mail: elbing@okstate.edu

# 1  Introduction

The Reynolds shear stress tensor in the Reynolds Averaged Navier-Stokes equations is indicative of the cross-gradient mixing process innate to turbulence, relaying that turbulence exhibits distinctive structural anatomy. This prompted several attempts to visualize such structure. It was not until the observations of Head & Bandyopadhyay,[1] which leveraged the physical insight of Theodorsen,[2] that provided the first convincing visual evidence of the near-wall three-dimensional (3D) vorticity structures; they termed such structures hairpin/horseshoe vortices. They detailed the evolution of these vortices influenced by the antagonistic events of stretching due to mean shear flow and the resulting induced velocities due to their intensified vorticity, causing them to lift-up in the wall-normal direction. Comparative studies of Brown and Thomas[3] and Deshpande et al.[4] would suggest that the observation of Head and Bandyopadhyay[1] were made for the dominant structure that developed through the turbulent boundary layer (TBL) at a 45° orientation. Such concrete observations led to a proposed model[5] that puts these 3D vorticity structures at the center of the momentum exchange process (i.e., they facilitate, *and potentially control*, the momentum exchange between the inner and outer regions (outer region here implies log region and beyond) of a TBL. A more recent investigation[6] revealed that such structures organize themselves to form long and meandering super-structures that are correlated spatially to produce regions of uniform momentum located in the TBL log-layer. Such super-structures have long been observed in moderately buoyancy-driven geophysical TBLs,[7] where the inner-outer coupling is tied to strong thermal updrafts. However, in the absence of buoyancy-driven thermals, as is the case in incompressible TBLs, understanding how near-surface coherent structures modulate outer layer dynamics is still a topic of active research with implications to flow control.

Given the centrality of such 3D vorticity structures to the characteristics of the near-wall TBL dynamics, it is natural to explore how these dominant coherent structures are modulated during polymer drag reduction (PDR) to explain the observations of Toms.[8] Several successful attempts[9-14] were pursued to this end. They found that polymer additives, in the near-wall region, increase spanwise spacing of the low-momentum streaks with a concomitant decrease and increase in the spatial frequency of the "burst" events and diameters of the vorticity structures, respectively, as drag reduction (DR) increases. Experimental observations[15] suggested a deficit in Reynolds shear stress for polymeric flows. This deficit was later linked computationally to polymeric stresses[16] as they engaged the pressure strain and diffusion terms of the Reynolds stress transport equation to enhance anisotropy of the fluctuating velocity field in the near-wall region.[17] This is in sharp contrast to turbulent flows over rough surfaces that exhibit an increase in drag and an accompanying decrease in fluctuating flow field anisotropy of near-wall turbulence.[18] Such developments suggest that polymer rheology is another key factor impacting PDR just as roughness scales impact drag enhancement.[19]

Polymers have a natural tendency to relax and maintain equilibrium conformations but tend to be stretched when in shear flow. However, to effect drag reduction, such relaxation times have to be greater than the local turbulent flow time scales. This effect is quantified by the Weissenberg number; a ratio of the polymer relaxation time to the flow time scales. Polymer stretching has been found to provide counter-torques that oppose the motion of hairpin/quasi-streamwise vortices (QSV), rendering them less intense.[20,21] This helps to explain the improved stability of the near-wall streaks,[20] inhibited $Q_2$ pumping of the hairpins, and, consequently, the damped Reynolds shear stresses.[21] The reduced swirling strengths of the vortices is also expected to alter the auto-generation cycle of the hairpins, assuming the modified cycle would have a critical strength threshold of the initial vortices required for its self-sustained continuation.[22] Flow simulations[23] have shown that this is the case with polymer molecules being actively engaged in re-routing the energetics of the flow to result in drag reduction.

Such a morphological introduction to PDR provides both fundamental and practical motivation for the current study. Much less is known about how the polymer additives modify the dynamics of dominant structures (that extend through the log region and beyond) and the corresponding flow statistics. A wide spectrum of structures in this region, together with increasing Reynolds number, make flow simulations for these regions computationally expensive. While earlier investigations[24-26] has shown that the mean velocity profiles in the log-layer deviate from the classical view[27] in the high drag reduction (HDR; > 40%) regime, a recent investigation[28] shows that statistically significant variation in log-layer slope occurs even at DR levels as low as 15%. Moreover, Elbing et al.[26] reviewed all experimental PDR studies within a TBL and showed that Reynolds number variation was insufficient to explain the deviations from the classical view in HDR, which indicates that polymer properties must play a role in HDR flow behavior. These developments show that characterization of the polymers is critical in analyzing modifications to the flow structures. Since many of the polymer properties are sensitive to the polymer concentration (e.g., Weissenberg number), the current study was performed with a developing TBL in a polymer ocean (uniform concentration) of polyethylene oxide (PEO).

Consequently, the current study experimentally measures a developing TBL velocity profile, with emphasis on the log-layer, that has been modified with the addition of drag reducing polymer solutions. The flow properties (freestream speed, downstream distance, viscosity, etc.) and polymer properties (concentration, mean molecular weight, etc.) are controlled such that various levels of DR are achieved with known Reynolds and Weissenberg numbers. Mean streamwise velocity profiles are presented, and then two-point correlations are used to investigate the modifications to the basic TBL structure, including the dominant structure inclination angles. The observations are compared with literature to assist with the interpretation of the PDR modified TBL structures. This includes assessing the effects of polymers on the auto-generation cycle of the hairpin structures. Ultimately, the current study aims to understand and provide insights on how polymer effects the coherency of induced motions as well as controls their spatial extent in the log-layer region and beyond. The remainder of the paper provides a description of the experimental methods in §2, §3 presents the results and discussion (3.1 Newtonian; 3.2 polymeric), and conclusions are summarized in §4.

# 2   Experimental Methods

## 2.1   Test facility and model

Testing was performed in the Oklahoma State University 6-inch low-turbulence, recirculating water tunnel.[29-31] The test section had acrylic walls for optical access and measured 1.1 m long with a 152 mm (6-inch) square cross-section. A 112 kW (150 hp) motor powered (MP44G3909, Baldor) a horizontal split case centrifugal pump (S10B12A-4, Patterson), and a variable-frequency-drive (EQ74150C, Teco) controlled the pump frequency to vary the tunnel speed. Flow conditioning (e.g. tandem configuration of honeycombs and settling chambers, 8.5:1 area contraction) resulted in an inlet turbulence level < 0.3% and negligible mean shear within the test-section core. Tunnel volume was determined by tracking changes in the conductivity of aqueous solutions of known amounts of added salts with a conductivity meter (CDH-287-KIT, Omega Engineering). The total volume ($1.47 \pm 0.04$ m$^3$) was determined from comparing the changes in conductivity against an established calibration curve.

A boundary layer trip (uniformly distributed 122 μm grit) at the test section inlet mitigated transitional effects on the tunnel walls. Measurements were acquired within the flat plate TBL that formed on the test section wall. Downstream of the boundary layer trip, the average surface roughness height ($R_a$) was at or below 0.8 μm. Converting $R_a$ to Colebrook type roughness ($k_c$)[32,33] and noting that the minimum viscous wall unit ($l_v$) was 7.5 μm, the maximum inner variable scaled roughness height ($k^+ = k_c/l_v$) in the current study was $k^+ \leq 0.5$. Thus, the surface was considered hydraulically smooth. The coordinate system used throughout the manuscript has the $x$ origin at the test section inlet and extending in the downstream direction, the $y$ coordinate increasing in the wall-normal direction with the origin at the test section centerline, and $z$ extends in the spanwise direction completing a right-handed coordinate system.

## 2.2 Instrumentation

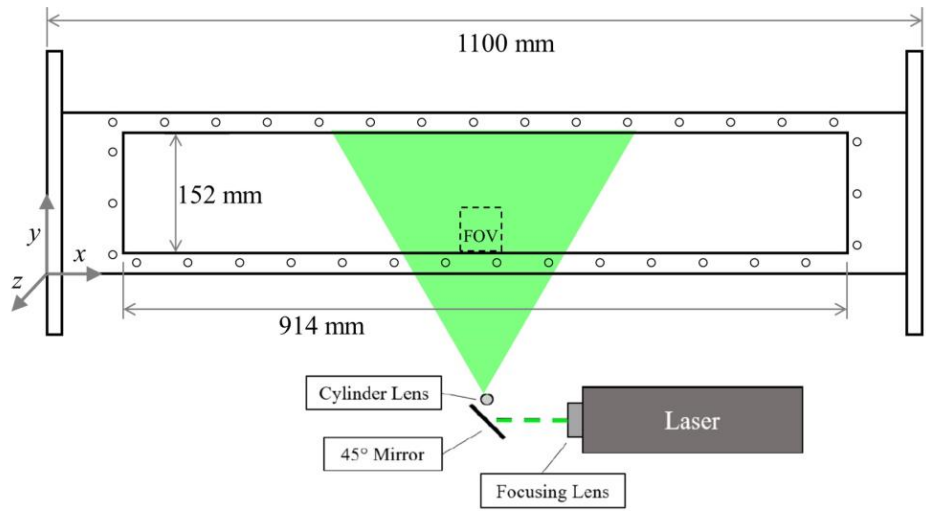

Figure 1. Schematic of the optical arrangement for the PIV measurements, including the nominal location of the FOV.

Time-resolved velocity vector fields were acquired at $x = 0.5$ m with two-dimensional particle image velocimetry (PIV). The image plane was aligned parallel to the flow, along the tunnel centerline ($z = 0$), and illuminated with a sheet of laser light. The laser sheet was formed from the beam of a high-speed Nd:YLF laser (DM30-527, Photonics) that was spread into a sheet with a cylindrical lens as illustrated in Figure 1. Hollow glass spheres (iM30K, 3M) with an average diameter of 18 μm were used to scatter the laser light. The scatter light was recorded at either 2.0 or 2.9 kHz with a high-speed camera (M110, Phantom) that had a resolution of 1280 pixel × 800 pixels. The camera was fitted with a 60 mm diameter, f/2.8D lens (AF Micro NIKKOR, Nikon) that resulted in a nominal field-of-view (FOV) of 10 mm × 15 mm and a spatial calibration of ~12 μm/pixel. The images were calibrated with a precision calibration target (Type 058-5, LaVision). The onboard camera memory allowed for 5000 images to be acquired for a single sequence. The PIV timing, acquisition, and image processing were performed using a commercial software package (Davis 8.2.3, LaVision). The images were processed using the standard multi-pass cross-correlation method with decreasing interrogation window sizes. The final window size was 32 pixels × 32 pixels with 75% overlap, which resulting in ~97 μm vector spacing. Note that for the primary test condition ($Re_\theta$=2000), the Newtonian viscous wall unit was ~14 μm, which sets the vector spacing at ≤ 6.9 viscous wall unit. The PIV uncertainty was quantified following the approach identified in Wieneke,[34] which uses the asymmetry in the correlation peaks when slightly shifted (~1 pixel) away from the optimized displacement to quantify the impact of image noise, including out-of-plane motion. For the current study the maximum uncertainty was ± 0.1 m/s (~ 0.3 pixels).

The water tunnel operation conditions (water temperature, static pressure, and pump motor frequency) were monitored throughout testing. The water temperature was measured with a T-Type thermocouple (TC-T-1/4NPT-U-72, Omega) located 0.92 m upstream of the contraction inlet. The static pressure was measured 76 mm upstream of the contraction inlet at the test section centerline elevation with a pressure transducer (PX230050DI, Omega). These measurements were recorded at 500 Hz throughout testing along with pump motor frequency from the variable-frequency-drive via a data acquisition card (USB-6218-BNC, NI) controlled with commercial software (LabView15.0.1, NI). The data acquisition system was also used for a subset of tests to record hot-film-anemometer (MiniCTA-54T42, Dantec) signals using a hot-film probe (55R15, Dantec) located at $x = 60$ mm. This was performed to confirm that the inlet turbulence intensity was below 1%.

## 2.3  Polymer preparation and characterization

The polymer used in the current study, PEO, has a structural unit (monomer) of (-O-CH$_2$-CH$_2$-) that results in a polymer backbone consisting of carbon-carbon (C-C) and carbon-oxygen (C-O) bonds. Polymer batches were prepared with manufacturer specified mean molecular weights ($M_w$) of $4\times10^6$ g/mol (WSR301, Dow chemical) or $8\times10^6$ g/mol (Sigma Aldrich) at concentrations between 1000 and 5000 ppm. The solutions were prepared via slowly sprinkling dry powder into a jet to prevent the formation of polymer aggregates. Trace amounts of sodium thiosulfate were used to mitigate polymer degradation due to background chlorine.[35,36]

Once fully hydrated, the polymer solutions were diluted to a desired concentration in the water tunnel and allowed to uniformly disperse via molecular diffusion. When fully mixed to produce a homogeneous polymer concentration (termed a polymer ocean), the water tunnel was operated at the test speed until a quasi-steady molecular weight, and therefore a nearly steady state molecular weight distribution, was achieved. Given the established knowledge that centrifugal pumps significantly degrade polymers; it was critical that the impact of flow assisted degradation on polymer flow properties be assessed. The results of this separate study are provided in Farsiani et al.,[37] which includes a detailed comparative analysis between mechanically degraded polymers within the polymer ocean and non-degraded samples at the same mean molecular weight. This showed that within the operation range of the current study, the degraded samples had the same drag reduction performance of non-degraded samples at the same mean molecular weight. It is also important to note, that polymer oceans were prepared at least 3 times and the molecular weight variation was within ~11%.

The mean molecular weights of the PEO polymer oceans were determined from the shear rate at the onset of drag reduction ($\gamma^*$), which is an established approach for estimating the mean molecular weight of PEO solutions.[38,39] Here the onset of drag reduction is determined from the intersection of the experimentally obtained drag reduction performance with the friction law for fully developed turbulent (smooth) pipe flow of a Newtonian fluid (i.e. the Prandtl-von Kármán law). Vanapalli et al.[38] compiled PEO data[27] to establish a relationship between $\gamma^*$ and the mean molecular weight, $\gamma^* = 3.35 \times 10^9 / M_w$. See Farsiani et al.[37] for a detailed description of the pressure drop apparatus used to acquire the polymeric pipe performance and Lander[40] for a detailed discussion of the measurement uncertainty. The overall uncertainty was determined from propagating the uncertainties associated with the temperature, fluid properties, pipe diameter, pressure drop per unit length (including uncertainty from pipe wall holes for the pressure taps[41]), and mass flowrate. While the uncertainty of a single condition varied significantly with the operation condition, the resulting uncertainty in the molecular weight was ~10%.[37,40] In addition, the pressure drop apparatus was used to determine the intrinsic concentration $[C]$ and intrinsic drag reduction $[DR]$.[42] The intrinsic polymer properties are used to establish a relationship between the polymer concentration ($C$) and the resulting drag reduction ($DR$), $DR/C = [C][DR]/([C] + C_M)$.[27,43,44] Choi & Jhon[44] showed that these intrinsic drag reduction properties for solutions of PEO-water were universal regardless of the flow geometry and solvent. Thus, the intrinsic properties could be used to determine the drag reduction relative to the Newtonian result within the polymer ocean.

## 2.4 Test conditions

Throughout the tests, the average water temperature was 22.0 ±1°C, which has a corresponding average kinematic water viscosity ($\nu$) of $9.55\times10^{-7}$ m$^2$/s. PIV was acquired at $x = 0.53$ m and at local freestream speeds ($U_\infty$) of 0.70, 1.68, and 3.22 m/s, which had corresponding nominal momentum thicknesses ($\theta$) of 1.2, 1.1, and 0.9 mm. The momentum thickness was determined by integrating the full velocity profile with a power-law curve fit, which the deviation from the exact velocity profile results in deviations of ~1% from DNS results at a comparable Reynolds number. Thus, the momentum thickness-based Reynolds number ($Re_\theta = U_\infty\theta/\nu$) ranged from 800 to 2900. The corresponding range of friction Reynolds number, $Re_\tau = u_\tau\delta/\nu$ (where $\delta$ is the 99% boundary layer thickness, $u_\tau = \sqrt{\tau_w/\rho}$ is the friction velocity, $\tau_w$ is the wall shear stress, and $\rho$ is the fluid density), for the three speeds were 350, 800 and 1540, respectively. For the Newtonian cases, flat-plate momentum integral analysis was used along with a traditional power-law curve fit of the scaled momentum thickness with the Reynolds number. This was previously shown for the current facility to result in a maximum deviation in $u_\tau$ of < 5%.[31]

For polymeric tests, the polymer ocean concentration was fixed at 100 ppm, and the steady-state mean molecular weights were varied from $0.7\times10^6$ to $4.2\times10^6$ g/mol. Polymeric test conditions were selected such that $Re_\theta$ was nearly constant at ~2000. As previously stated, the wall shear stress and friction velocity were determined from the corresponding Newtonian condition and the intrinsic drag reduction properties of PEO.[42,44] This resulted in DR levels from 20% to 62% and Weissenberg numbers ($We = \lambda u_\tau/l_\nu$, where $\lambda$ is the polymer relaxation time and $l_\nu = \nu/u_\tau$ is the viscous wall unit) from 0.22 to 3.24, in addition to the water condition (DR = 0%). For dilute polymer solutions (i.e. separation between individual polymer chains is sufficiently large that there is minimal chain-to-chain interaction in solution), the relaxation time can be set equal to the Zimm time,[45]

$$\lambda_z = 0.422 \frac{[\eta]_o \mu_S}{RT} M_w,$$

where $R$ is the ideal gas constant, $T$ is the absolute temperature, and $[\eta]_o$ is the intrinsic viscosity. The intrinsic viscosity was estimated from the Mark-Houwink relationship[46]

$$[\eta]_o = 0.0125 M_w^{0.78}.$$

Polymer solutions are considered dilute when the concentration is below the overlap concentration, $C^* = 1/[\eta]_o$, which for the current study $C^* \geq 517$ ppm. Thus, all test conditions in the current study are considered dilute and, consequently, only depend on the molecular weight. Other critical non-dimensional polymer dependent parameters include the ratio of the solvent viscosity to the zero-shear viscosity of the polymer solution ($\mu^*$) and the length ratio of the fully extended to coiled polymer molecules ($L$). These parameters had minimal variation in the current study with the ranges of $\mu^*$ and $L$ being 0.93 to 0.99 and 125 to 308, respectively.

## 2.5  Data analysis

The spatial coherence and the statistical significance of various features within the velocity field were quantified using two-point spatial correlations ($R_{uu}$) of the streamwise velocity fluctuations. The functional form of $R_{uu}$ is

$$R_{uu}(x_{ref}, y_{ref}, \Delta x, \Delta y) = \frac{\langle u(x_{ref}, y_{ref}) u(x_{ref} + \Delta x, y_{ref} + \Delta y) \rangle}{\sigma_u(x_{ref}, y_{ref}) \sigma_u(x_{ref} + \Delta x, y_{ref} + \Delta y)}. \tag{1}$$

Here, $x_{ref}$, $y_{ref}$ are the spatial location of the reference point, $\sigma_u$ is the standard deviation of $u$ at the specified location, $\langle \; \rangle$-brackets indicate an ensemble average, and $\Delta x$, $\Delta y$ are the streamwise and vertical separation distances from the reference location, respectively. For a true ensemble average $R_{uu}$ is a function of $x_{ref}$, $y_{ref}$. For practical reasons, a surrogate ensemble average, $\langle x \rangle = \frac{1}{N}\sum_{i=1}^{N} x_i$ with $N = 4000$ PIV vector fields is adopted. Here ($N$) was selected based on the convergence of the correlation curves, which for DR = 62%, had the data converged by $N \sim 3000$ with uncertainty < 5%. The number of points in the ensemble average decreases linearly with increasing separation distance due to the finite-sized FOV, which Taylor's frozen turbulence hypothesis was invoked to mitigate this issue.

Inclined motions frequently results in elliptically shaped two-point correlation maps with the principal axis inclined at an angle away from the wall, termed the structure inclination angle. Their identification follows the general analysis of Marusic.[47] Following this, the dominant structure inclination angle (α) can be found from the slope, $\alpha = \tan^{-1}\left(\langle \Delta y / \Delta x_{pk} \rangle\right)$, where $(\Delta x_{pk})$ is the streamwise separation distance of the peak in the two-point correlation.

# 3   Results and Discussion

## 3.1   *Newtonian (Baseline) Results*

### 3.1.1   Newtonian mean velocity profiles

The inner variable scaled mean streamwise velocity $(U^+ = U/u_\tau)$ profiles for water (Newtonian, DR = 0%) at three different $Re_\theta$ values as shown in Figure 2, where $y^+ (= y/l_\nu)$ is the inner-variable scaled wall-normal distance. The current results are compared with DNS data[48] at a similar Reynolds number as well as against the traditional law of the wall $(U^+ = \ln(y^+)/\kappa + B)$ with the von Kármán coefficient (κ) and slope intercept (B) equal to 0.4 and 5.0, respectively. At this Reynolds number range, the log-region is relatively thin and consequently difficult to identify. Thus an indicator function, $\zeta = y^+ \, dU^+/dy^+$, was used to identify the log layer region of the TBL (i.e. region of constant ζ). This examination showed that the log layer spanned 80 < y$^+$ <230 for the $Re_\theta$ = 2000 condition, which provides guidance on locations of points being correlated in the subsequent analysis. Note that this range is consistent with the conclusions of Marusic et al.[49] that used the conservative bounds of the log-region to be $3Re_\tau^{0.5} < y^+ < 0.15Re_\tau$ from very high Reynolds number experiments.[50-53] This requires $Re_\tau > 40,000$ to have one decade of log-region, and for the current condition it would place the log-region spanning $84 < y^+ < 120$. Since the bounds are on the same order of magnitude, it is debatable if there is a true log-region. However, for the current discussion and analysis the "log-region" refers to that identified from the indicator function, $80 < y^+ < 230$.

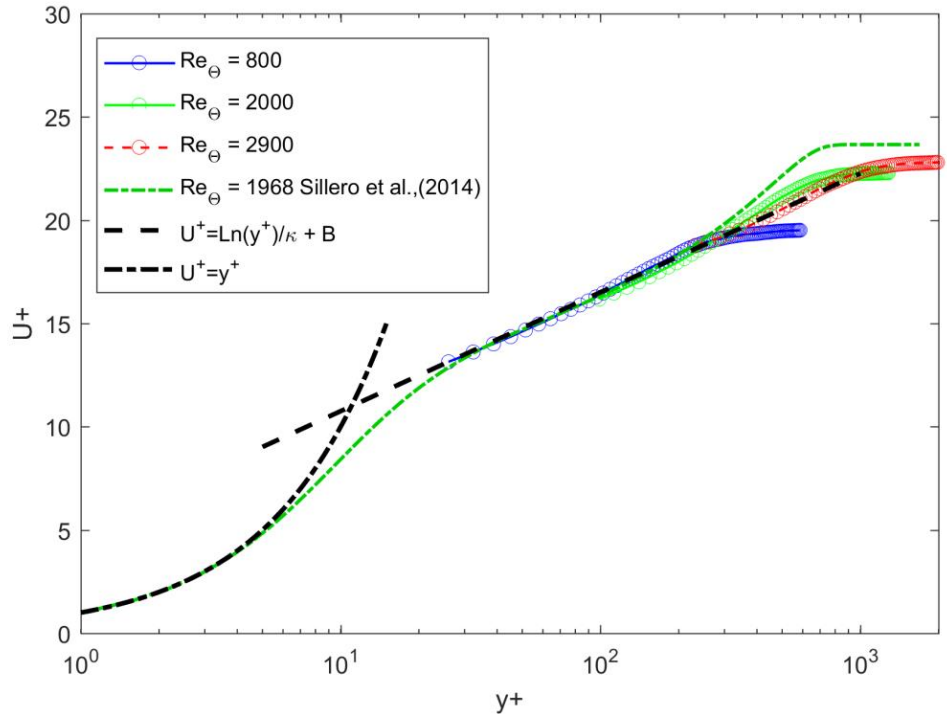

Figure 2. Inner variable scaled streamwise velocity profiles for water (Newtonian, DR = 0%). The profiles are compared with the viscous sublayer profile ($U^+ = y^+$) and the traditional log-law profile ($U^+ = \ln(y^+)/\kappa + B$) with $\kappa = 0.4$ and $B = 5.0$. DNS data (Sillero et al.[48]) at $Re_\theta = 1968$ is included for reference.

### 3.1.2 Newtonian two-point correlations

The reference location for the two-point correlations shown in Figure 3 were fixed within the log-region ($y_{ref}^+ = 148$), though on the outer edge for the lowest Reynolds number. The correlations from the three Reynolds numbers ($Re_\theta = 800$, 2000, and 2900) are shown with the streamwise separation length scaled with $\delta$. The same overall trend is observed for all three Reynolds numbers; the peaks decrease with increasing wall-normal separation ($\Delta y$), the streamwise separation ($\Delta x$) for the peak correlation increases with increasing $\Delta y$, and the correlations are asymmetric about $\Delta x = 0$. Coherent motions with streamwise aligned orientation would suggest upstream-downstream asymmetry about $\Delta x = 0$ (zero streamwise separation) and progressively decreasing peak value of correlations with increasing wall-normal separation ($\Delta y$) along the angle of their inclination, as observed in Figure 3. In the current study, the dominant structure inclination angle found from the peak locations with various separation distances was $46.9° \pm 4°$. It is important to note here that the mean structure angle within the log-region is well established in the literature[3,4,54] to be between $15°$ and $18°$. However, when an isolated dominant structure has been considered, their inclination angle has been found to be ~$45°$.[1,4] Given that the region correlated in the current study extends from middle of log region to the outer region, this suggests that the measured angle corresponds to the dominant structure extending through the log layer to the outer regions.[4] The subsequent polymeric analysis compares the deviations of these dominant structures with increasing drag reduction.

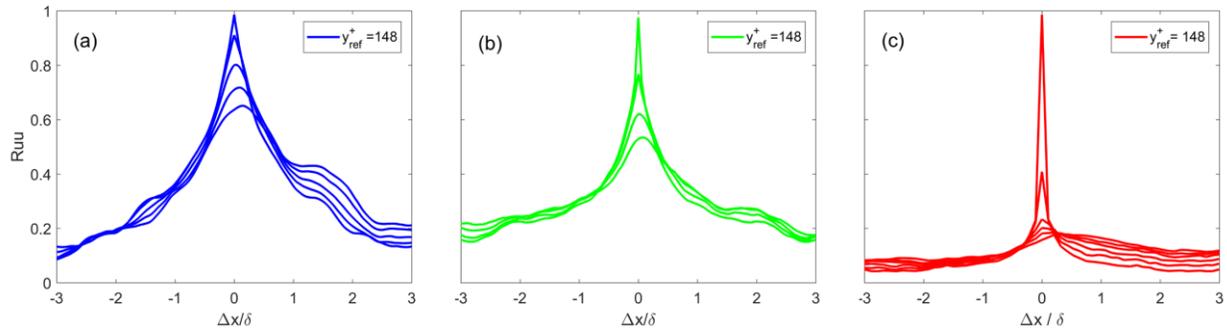

Figure 3. The two-point correlations at a fixed inner variable scaled wall-normal reference location ($y^+_{ref} = 148$) with maximum $\Delta y^+ = 100$ for $Re_\theta$ of (a) 800, (b) 2000, and (c) 2900. Note that each of the consecutive lines show a spatial spacing of $y^+ = 20$.

In Figure 4, the variational traits of the flow correlations in purely streamwise direction ($\Delta y^+ = 0$) are compared with correlations with $\Delta y^+= 100$ for three different reference heights. The reference heights have been kept the same for the two cases. Figure 4a depicts the correlation of motions at the lower and upper bounds of log region as well as the correlation of log layer events with that in the outer region. The decrease in the correlation peak quality factor in Figure 4a indicates that coherence in these motions is higher and persist over enhanced streamwise lengths in the outer log region of the TBL. This reinforces the idea of concatenated formation of hairpins in a packet like organization to produce such strong flow patterns in the outer log region and beyond.[55] Moreover, peaks have a slight positive offset about $\Delta x = 0$ in Figure 4a, showing that the responsible structures for these correlations lean downstream. Figure 4b shows that the correlations progressively increase for $\Delta x > \delta$ as upper portions of the log layer are realized. The dependence of $R_{uu}(\Delta x, 0)$ on the reference height shows a slower rate of decay in the correlation magnitude with streamwise separation relative to Figure 4a and is consistent with Monty et al.[56] Except, in this case the systematic dependence of $R_{uu}(\Delta x, 0)$ on $y_{ref}$ is relatively subdued as compared to $R_{uu}(\Delta x, \Delta y)$, and the curves are increasingly similar for chosen $y_{ref}^+$ values that are closer to the upper bound of the log layer.

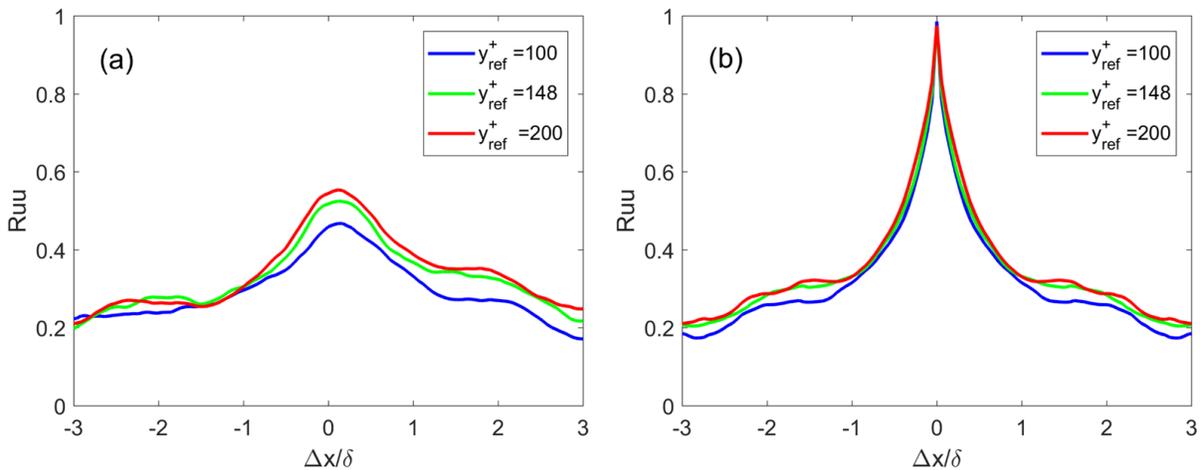

Figure 4. Two-point correlations of the streamwise fluctuating velocities in water (Newtonian) at $Re_\theta = 2000$ for varying reference heights and (a) $\Delta y^+ = 100$ and (b) $\Delta y^+ = 0$.

## 3.2  Polymeric Results

### 3.2.1  Mean velocity profiles

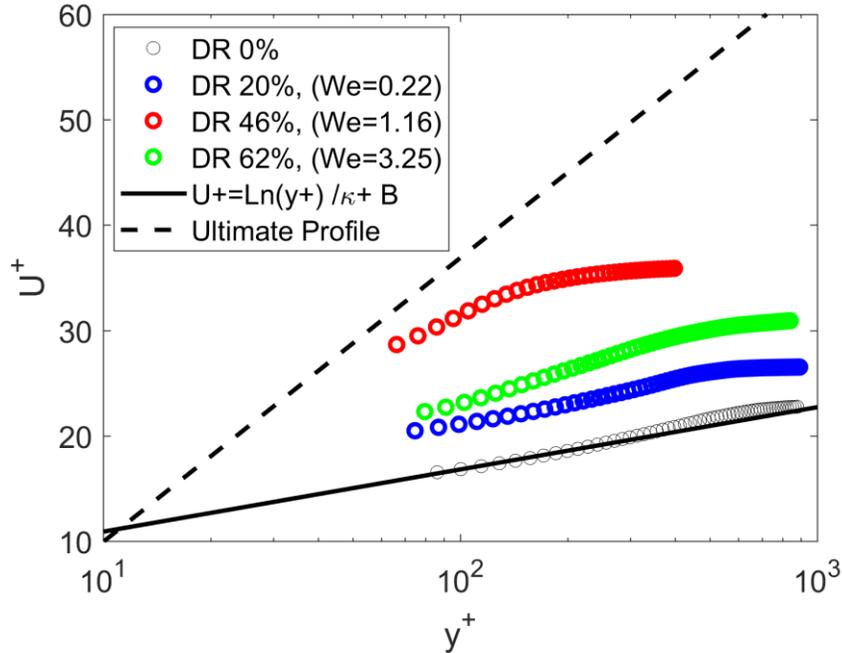

Figure 5. Inner variable scaled mean streamwise velocity profiles at three DR levels (20%, 46%, and 62%) with their corresponding *We* listed. The polymer ocean concentration was 100 ppm and $Re_\theta = 2000$. The log-law, ultimate profile, and water results (DR = 0%) are included for comparison.

The inner variable scaled mean streamwise velocity profiles for the polymer oceans at a concentration of 100 ppm are shown in Figure 5. Drag reductions of 20%, 46%, and 62% were produced with quasi-steady mean molecular weights of 0.7, 2.2, and 4.2 $\times 10^6$ g/mol, respectively. Included in the figure for reference are the traditional law of the wall ($\kappa = 0.40$, B = 5.0) and the ultimate profile ($U^+ = 11.7 \ln(y^+) - 17.0$).[43] The results show the well-established[15,20,25-27,39,57,58] trend of an upward shift in the log-layer that is proportional to the drag reduction level. These mean profiles are used primarily to identify the log-layer for the subsequent analysis. However, the behavior within the log-layer is an open research question,[25,26] particularly at HDR (DR > 40%). The classical view[27] is that only the slope increment (*B*) increases with increasing DR, but recent work[25,26,28] has shown that the von Kármán coefficient ($\kappa$) is dependent on both the polymer properties and Reynolds number, especially at HDR. The Weissenberg number (*We*) is provided for each condition in Figure 5 as it is a likely parameter governing this dependence. Figure 5 also demonstrates the condensed log-layer since the extent of the buffer layer increases with increasing DR,[19,58,59] especially at HDR. Note that the mean velocity profiles at additional DR levels (not shown) were measured to confirm this general trend. Overall, the mean profiles indicate that the polymer dynamics and properties play a key role in controlling the flow structure.

### 3.2.2 Polymeric two-point correlations

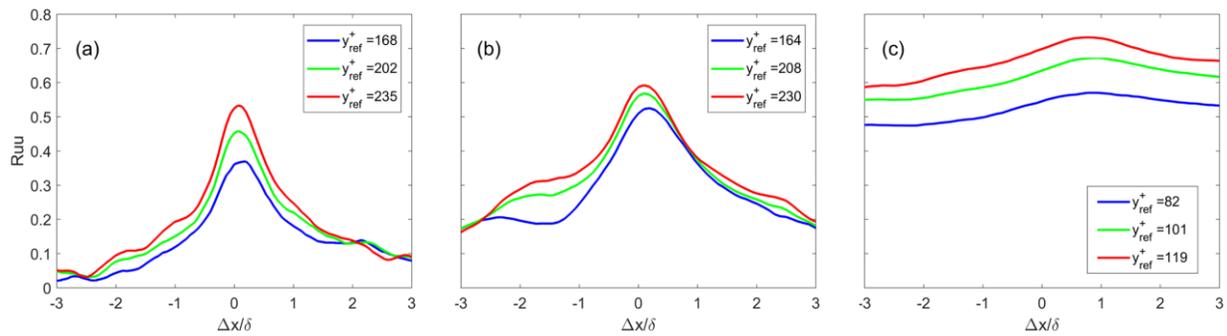

Figure 6. Two-point correlations with a fixed wall-normal separation ($\Delta y^+ = 100$) and $y_{ref}$ within the log-layer $\left(82 \leq y_{ref}^+ \leq 235\right)$ at drag reduction levels of (a) 20%, (b) 46% and (c) 62%.

There is a general consensus that polymers significantly enhance anisotropy of the turbulent fluctuations in the near-wall region.[15,17] Given this streamwise bias of flow statistics, Figure 6 shows the streamwise scale of the flow structures in the log-region within the low drag reduction (LDR) regime (DR=20%) the boundary between LDR and HDR (DR = 46%) and within the HDR regime (DR= 62%). Here the wall-normal separation was fixed at $\Delta y^+ \cong 100$ with the reference locations selected to relate the flow statistics within the upper half of the log-layer and potentially extending to the outer-wake region for all the DR regimes. Comparison between the LDR regime (Figure 6a) and that of water (DR = 0%; Figure 4a) shows minimal variation. However, the differences relative to water progressively increases moving from LDR (Figure 6a) to the LDR/HDR boundary (Figure 6b) and then to HDR (Figure 6c). The same general trend is observed as DR levels increase; their peak values increase while the peak quality factors (sharpness) decrease as upper regions of the TBL are correlated. This is particularly apparent for DR = 46%, where not only the streamwise-spatially averaged $R_{uu}$ increases with $y_{ref}$, indicative of increase in the span of uniform momentum zones, but the different curves also show a tendency to converge at streamwise separation distances closer to the peak ($\Delta x/\delta \sim 0.7$).

The nuances in the trends observed for DR = 46% extend to the DR = 62% case, but there are significant deviations in the trends indicating that the polymer additives are significantly influencing the structure within this region of the TBL. These deviations potentially arise from the modulations in the viscous length scale ($l_\nu$) due to the polymer additives as well as outer boundary layer scales. As shown in Figure 5 for HDR, the log-layer is significantly reduced, and consequently the coherence between the inner and outer regions of the TBL are enhanced as seen by strong velocity correlations in Figure 6c. Figure 6c shows a significant increase in $R_{uu}$ with the increased correlations persisting over a significantly larger streamwise scale, which suggests that the flow becomes more streamwise dominated with increasing DR.[15] Also, note that the variation in magnitude of the correlations with streamwise separation occurs over a relatively narrow range as the outer regions (higher $y_{ref}$) are approached. This remarkable *stabilization* of the streamwise correlations points to the possible existence of unusual prolongations in the streamwise direction of the uniform momentum zones, which were originally reported to exist in Newtonian flows.[6] To confirm this, one needs visualizations of the flow field over a sufficiently large spanwise-streamwise plane that is currently unavailable. While the streamwise fluctuating velocity correlations increase noticeably in the streamwise direction, it is important to also note its increase along the wall-normal direction (i.e. the spatial scale increases in both the streamwise and wall-normal directions, though more rapidly in the streamwise direction).

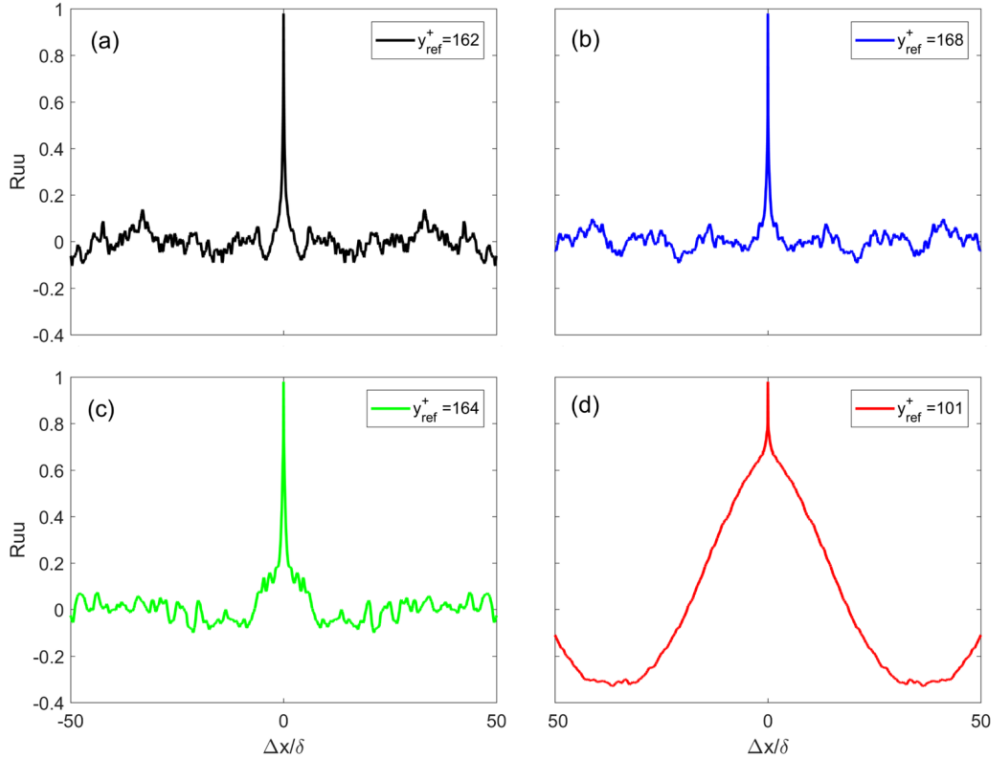

Figure 7. Two-point correlations of the streamwise velocity fluctuations with $\Delta y^+ = 0$ and $101 \leq y_{ref}^+ \leq 168$ with (a) DR = 0%, (b) DR = 20%, (c) DR = 46%, and (d) DR = 62%.

Figure 7 shows the correlations of the streamwise fluctuating velocities with purely streamwise separations (i.e. $\Delta y = 0$), which reveals additional details of how the polymers modify the flow statistics. Note that for clarity only one $y_{ref}$ was selected for each DR level in Figure 7, and the selections were such that it depicts the events within the log-layer. Again, the deviations in the flow statistical trends are mild between water (DR = 0%; Figure 7a) and the LDR regime (Figure 7b), but as the DR level increases the differences become more apparent. In particular, note that the correlations persist longer spatially in the streamwise direction as DR increases. While the spatially averaged values of the normalized correlations increase with increasing DR, the fluctuations in the correlations are progressively reduced. This strongly suggests that the flow *stabilization* is spatially more prevalent with increasing DR. The extent of such *laminarization* expands in the *x-y* plane as a dominating effect of the polymer additives, which has been well-known to restructure the flow energetics.[16,21,60]

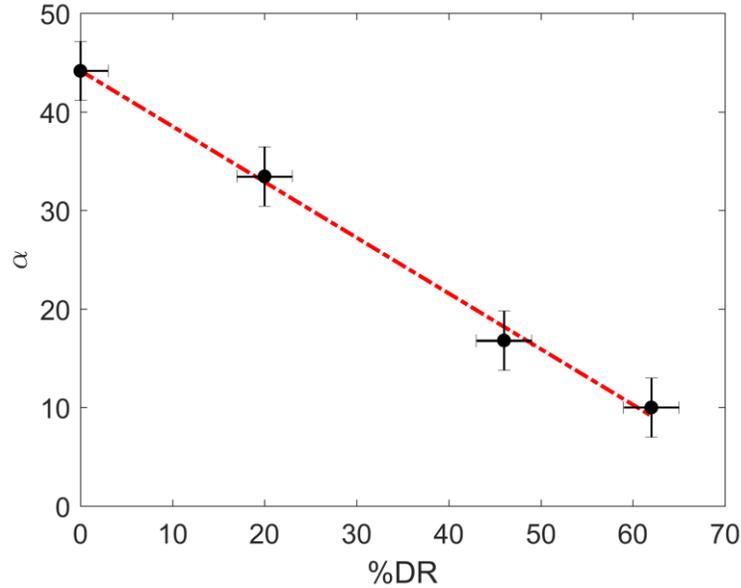

Figure 8. Dominant structure inclination angles within the log-layer plotted versus DR level. Data from polymer oceans at a concentration of 100 ppm with molecular weight varied and $Re_\theta = 2000$. Dashed line is the linear best-fit curve to the data.

As noted in the discussion for Figure 6, previous findings[15,17,56,61] documented that the polymer additives make the flow more parallel with the streamwise direction with increasing DR. Consequently, the dominant flow structure inclination angle within the log-layer is plotted versus drag reduction in Figure 8. This shows that the maximum inclination angle of the dominant structure within the log-layer (~45°) occurs with DR = 0% (water), and the minimum inclination angle (~10°) occurs with DR = 62%. Linear regression analysis shows that the slope and intercept were statistically significant ($p < 0.05$) with the slope and intercept being −0.563 ± 0.085 deg/% and 46.9° ± 3.2°, respectively, with 95% confidence. Extrapolation of this linear-fit to a zero inclination angle results in a maximum drag reduction of 77%. This suggests that the existence and the ensuing spatio-temporal evolution of vortices represent a major chunk of events in near-wall turbulence that are responsible for energy exchange between the mean and turbulent flows and therefore a significant rise in skin friction drag as compared to laminar flows, which lacks both in the former process and the latter effect. Such vortical interactions have been found to be an integral part of the autonomous cycle of near wall turbulence.[62] The flattened orientation of the near-wall structures means significant bias of the flow scales in the stream-wise direction with increasing DR. One immediate inference from the observed anisotropy is that these structures tend to rise less through the TBL. This would suggest that the velocity induction effects, responsible for lifting these structures through the TBL,[1] have been curtailed as they are weakened by the presence of the polymers.[16,20] This is consistent with the DNS findings of Kim et al.[21] that suggests that polymers provide torques opposing the motions of the vortices, and thereby, reducing their ability to influence events associated with their strength. Dubief *et al.*[20] proposed a mechanistic explanation that stability of the momentum streaks was enhanced at the expense of weakened vortex structures; polymers stretch as they wrap around the vortices, extracting energy and re-injecting it into the near-wall ($y^+$ > 5) momentum streaks. The wrapping action is consistent with polymers providing torques, opposing vorticity intensification. If this were to be true, the structure inclination angles would decrease, as observed in Figure 8. This makes their contributions to the flow statistics more in the streamwise direction rather than the wall-normal direction (i.e. reduces their inclination angle), thereby reducing the momentum transfer between the outer and inner regions of the TBL.

## 4   Summary and Conclusions

Modification of the dominant coherent structures in the polymeric drag reduced TBLs were investigated by examining the flow statistics near the wall. Coherent motions/structures are 3D regions of flow exhibiting significant correlations between flow parameters.[63] This definition was incorporated to infer orientations of dominant structures within the TBL log-region by computing two-point normalized correlations of streamwise velocity fluctuations. Results for both the Newtonian and polymeric flows were obtained, compared, and analyzed. Mean velocity profiles identified the extent of the log-regions and compared well with DNS data at a similar Reynolds number.[48] The Newtonian results were consistent with available literature[1,4] with the dominant structure inclination angle being $46.9° \pm 4°$ (average from $Re_\theta = 800, 2000,$ and $2900$) though it should be noted that the *mean* structure angle within the log-region is between $15°$ and $18°$.[3,4,54] Log-regions farther from the wall showed progressively improved correlations. This indicates that the uniformity of momentum zones has expanded in *x-y* plane.

Trends in the two-point correlations for the polymeric flows deviated from the Newtonian flows, particularly at HDR (>40%). Much like that of the fluctuating velocity fields in the buffer and viscous sublayer, flow scale anisotropy was progressively enhanced in the upper bounds of the log and outer-wake regions; structures have significantly enhanced streamwise scales. This anisotropy is particularly prominent for regimes of DR close to MDR. Correlations show the reduced range of variations for a given streamwise scale as DR increases, suggesting the stabilizing effect of polymer additives on the flow statistics. For the same extent of the TBL, the correlations in polymeric flows were found to be marginally higher than the Newtonian flows for LDR. These trends were then found to deviate appreciably for HDR. The details of how the momentum zones (likely due to LSM) tend to develop as Weissenberg numbers increase, is inferred from the normalized correlation plots at three DR levels (20%, 46%, and 62%).

Normalized correlations purely along the streamwise direction ($\Delta y = 0$) corroborate the stretching of streamwise scales of the flow with increasing DR, but also show suppression in the fluctuations of the correlations. This indicates that polymer additives are actively engaged in suppressing the fluctuating content of the flow structures at higher frequencies. This serves as an experimental verification of the flow simulations[23] featuring weakened vortices, due to opposing torques by polymer body forces.[21] Such vortices would then contribute less to the auto-generation process of hairpins.[22] Such attenuation of the fluctuating turbulent motions would suggest that the offspring shedding process of the primary hairpins (flow structures) has been curbed and that this restriction becomes stronger in HDR flows. However, modes of action on these structures by polymers is rather selective, based on the intrinsic polymer and flow properties. Swirling-strength based time scales of the vortices suggest that the shorter the time scale, the stronger the structure. This would increase the local Weissenberg numbers and therefore the probability of being attenuated by the stretching action of the polymers. Since the most intense of these vortices are in the buffer-layer, in the LDR the effect of polymers would remain primarily within the buffer-layer. This would explain the mild transition in observation between DR = 0% and DR = 20% for all correlations in the log-region. Mild transitions, however, do not suggest that there is "virtually" no effect of polymers additives on the log-layer. It is clear that the "domino effect" of the weakened hairpin/quasi-streamwise vortices or QSVs is to make the log-region structures more inclined towards the flow direction. But as higher Weissenberg numbers are achieved, the polymers start effecting the less energetic vortex regions (heads and necks) found in the log-layer, directly. This could be attributed to the polymer relaxation times being significantly larger than the largest time scales associated with the vortex structures of the TBL. The reduced offspring shedding activity of the vortices, can potentially explain the enhanced correlations observed over long streamwise scales (Figure 6 and Figure 7). The fact that the structures are far apart, is compensated by their flattened orientation. Being targeted by the polymers based on their strengths, vortex structures with similar strength could weaken simultaneously, which aids in preserving the coherence of the dominant eddy structure.

Figure 8 shows the dependence of the structure inclination angle on DR. Decreasing averaged inclination angle of coherent structures with the streamwise direction show that anisotropy of the flow scales increases in proportion to the reduction in skin friction and thus reducing momentum transfer between the outer and inner regions of the TBL. Weakened vortices near the wall (log-region) due to the presence of polymers bias their contribution to the flow statistics towards the streamwise direction.[15] Moreover, the flattened inclinations, shown in Figure 8, suggest a linear relationship between such structure angles (in the TBL log-region) with increasing DR, $\alpha = -0.563DR + 46.9$. Extrapolation of this fit predicts a maximum drag reduction of 77%. This further suggests that polymers are involved in severely mitigating the spatio-temporal evolution of vortical structure in the TBL near wall region, a process which significantly contributes to the production of skin-friction drag.

## Acknowledgments

This work was sponsored in part by the National Science Foundation (NSF) under grant 1604978 (Ronald Joslin, Program Manager). Jayaraman was partially supported by NASA OK-EPSCoR and Baker-Hughes General Electric.